\documentclass[conference]{IEEEtran}

\usepackage{blindtext, graphicx}
\usepackage[dvipsnames]{xcolor}
\usepackage{cite}

\usepackage[load-configurations=binary,detect-all,binary-units=true,range-phrase=--,per-mode=symbol]{siunitx}
\usepackage{acronym}
\usepackage{amsmath}

\usepackage{amsthm}
\usepackage[capitalise]{cleveref}
\usepackage[caption=false,font=footnotesize]{subfig}
\usepackage{multirow}
\usepackage{tabularx}
\usepackage{array}
\newcolumntype{P}[1]{>{\centering\arraybackslash}p{#1}}
\usepackage{resizegather}
\usepackage[utf8x]{inputenc}
\usepackage[T1]{fontenc}
\usepackage{textcomp}
\usepackage{balance}

\usepackage{makecell} 
\setcellgapes{5pt}

\acrodef{CNN}{Convolutional Neural Network}
\acrodef{PLP}{Perceptual Linear Predictive}
\acrodef{AI}{Artificial Intelligence}

\begin{document}
	
	\title{Can Machine Learning Be Used to Recognize and Diagnose Coughs?}
	
	\author{
		\IEEEauthorblockN{Charles Bales\IEEEauthorrefmark{1}, Muhammad Nabeel\IEEEauthorrefmark{1}, Charles N. John\IEEEauthorrefmark{1}, Usama Masood\IEEEauthorrefmark{1}, Haneya N. Qureshi\IEEEauthorrefmark{1}, \\ Hasan Farooq\IEEEauthorrefmark{1}, Iryna Posokhova\IEEEauthorrefmark{2}\IEEEauthorrefmark{3} and Ali Imran\IEEEauthorrefmark{1}\IEEEauthorrefmark{3}}
		\IEEEauthorblockA{\IEEEauthorrefmark{1}AI4Networks Research Center, Dept. of Electrical \& Computer Engineering, University of Oklahoma, Tulsa, USA}
		\IEEEauthorblockA{\IEEEauthorrefmark{2}Kharkiv National Medical University, Kharkiv, Ukraine}
		\IEEEauthorblockA{\IEEEauthorrefmark{3}AI4Lyf LLC, USA}
		\parbox{\linewidth}{\centering
			\texttt{csbales@wpi.edu, \{muhmd.nabeel,charlesj,usama.masood,haneya,hasan.farooq\}@ou.edu,\\iryna@ai4lyf.com, ali.imran@ou.edu}}
	}%

	
	\maketitle
	
	\begin{abstract}
		Emerging wireless technologies, such as 5G and beyond, are bringing new use cases to the forefront, one of the most prominent being machine learning empowered health care. One of the notable modern medical concerns that impose an immense worldwide health burden are respiratory infections. Since cough is an essential symptom of many respiratory infections, an automated system to screen for respiratory diseases based on raw cough data would have a multitude of beneficial research and medical applications. In literature, machine learning has already been successfully used to detect cough events in controlled environments. In this paper, we present a low complexity, automated recognition and diagnostic tool for screening respiratory infections that utilizes \acp{CNN} to detect cough within environment audio and diagnose three potential illnesses (i.e., bronchitis, bronchiolitis and pertussis) based on their unique cough audio features. Both proposed detection and diagnosis models achieve an accuracy of over \SI{89}{\percent}, while also remaining computationally efficient. Results show that the proposed system is successfully able to detect and separate cough events from background noise. Moreover, the proposed single diagnosis model is capable of distinguishing between different illnesses without the need of separate models.
	\end{abstract}
	
	
	\acresetall
	
	%
	
	\section{Introduction}
	
	A study conducted in 2016 estimated that \SI{4.4}{\percent} of all deaths in that year, a number surpassing 2 million, were due to various lower respiratory tract infections in both young children and adults~\cite{Troeger2018}.
	To prevent this in the future, one option is to come up with methods that could perform early detection of potential respiratory tract infections so that the likelihood of severe complications at a later time could be reduced.
	In most of the respiratory infections, cough is an important and essential early symptom~\cite{irwin2018classification}.
	This means that cough can be used for early detection of these respiratory conditions.
	However, when a person starts coughing, it is nearly impossible for a common person to diagnose the underlying disease at home.
	Unfortunately, when the infected person decides to visit a hospital to perform clinical testing, it may already be too late.
	Thus, an automated system for the detection and preliminary diagnosis of respiratory infections based on cough events would serve as a useful tool in the medical field.
	
	In the literature, it has been shown that it is possible to use traditional signal processing techniques to process and recognize different cough sounds~\cite{al2013signal}.
	This is  done with the help of unique latent features  in  cough  sounds.
	To have more robust and cost-effective solutions, researchers are now turning their focus towards \ac{AI} techniques.
	In recent years, many \ac{AI} based systems have been developed to either detect and separate cough sounds from different types of environmental sounds~\cite{Amoh2016} or to preliminary diagnose different respiratory conditions from prerecorded cough sounds~\cite{porter2019prospective}.
	However, most of these approaches involve either a heavy pre-processing phase or inherent design restrictions.
	This includes discarding frames of silent audio through the use of an RMS energy threshold to only categorize selected audio~\cite{Amoh2016}, specifically selecting noise-reducing hardware for data collection~\cite{Aykanat2017} and utilizing complex hardware~\cite{mlynczak2015automatic}.
	Moreover, the categorization of audio inputs as different types of coughs requires a more complex machine learning model, making maintaining usefulness for mobile applications more difficult.
	This is because the minimization of processing power has been of less priority as compared to the model performance.
	
	\begin{figure*}[t]
		\centering
		\includegraphics[width=\linewidth]{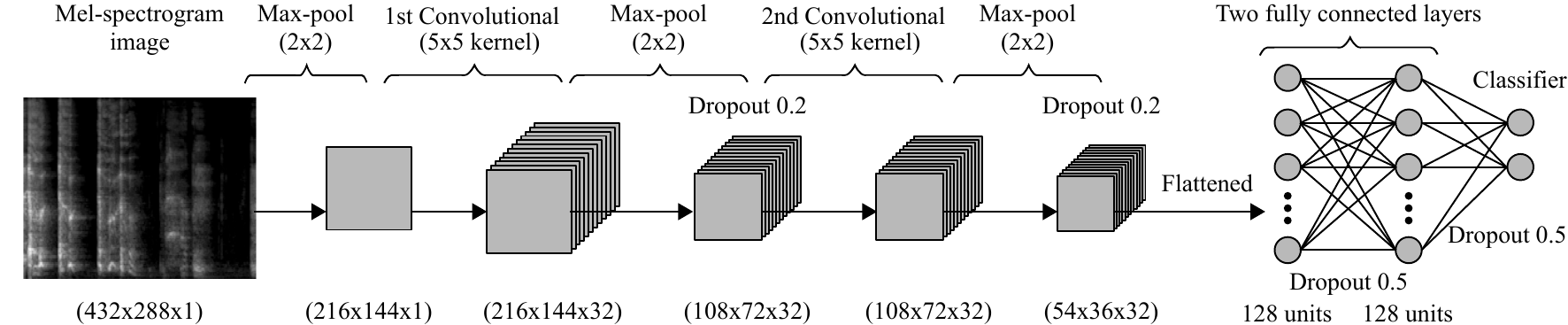}
		\caption{Overview of the CNN structure used for cough detection.}
		\label{fig:cnn}
		\vspace{-12pt}
	\end{figure*}
	
	In the literature, different types of audio categorization has been well investigated and among the various machine learning based techniques, \ac{CNN} based approaches are a popular choice for this task~\cite{kim2019comparison}.
	Using three publicly available datasets, ESC-10, ESC-50, and UrbanSound8K, that contain a large collection of environmental sounds, it is shown that a \ac{CNN} model can successfully be applied to environmental sound classification~\cite{piczak2015environmental}.
	The success of the \ac{CNN} based approach in categorizing audio into the fifty categories of the ESC-50 set, one of these being cough sounds, suggests that the network model may be capable of doing so with just coughs specifically.
	Indeed, this approach has been attempted in several other recent works~\cite{Amoh2016,Aykanat2017}.
	
	For diagnosis of specific illnesses via cough audio, authors in~\cite{Pramono2016} proposed an algorithm capable of identifying pertussis accurately, whereas in~\cite{Sharan2019}, authors exploited several machine learning models to diagnose croup using features extracted from cough audio.
	Results of similar other studies, e.g.,~\cite{porter2019prospective}, show that coughs originating from specific infections or illnesses have a sufficient number of distinguishing features that machine learning models can use for categorization.
	Taking the next step of sorting through multiple potential causes simultaneously is what our work seeks to achieve.
	
	Specifically, in this work, we use \acp{CNN} to first detect and separate cough sounds from different types of sounds.
	We then use the detected cough sounds to diagnose three potential illnesses (i.e., bronchitis, bronchiolitis and pertussis) based on their unique cough audio features in a unified framework.
	This is contrary to prior works, e.g.,~\cite{Sharan2019,Pramono2016}, that use a binary-class model to diagnose a particular disease (i.e., categorizing among that particular disease and no disease class).
	Also the main idea in our work is to significantly minimize recorded audio pre-processing before feeding data into a network model while maintaining acceptably high levels of accuracy for cough detection, among other metrics. 
	
	Our main contributions can be summarized as follows:
	
	\begin{itemize}
		\item We propose a low-complexity \ac{CNN} model for the detection of cough events from a typical audio segment with environmental noise.
		\item We then categorize among three common respiratory conditions (i.e., bronchitis, bronchiolitis and pertussis) with cough symptom in a unified work.
		\item The dataset used to train our proposed model does not go through the typical level of pre-processing that other works use for cough detection or diagnosis.
	\end{itemize}

	\begin{table*}
		\caption{Performance Metrics for Cough Detection.}
		\centering
		\begin{tabular}
			{|P{2cm}|P{2cm}|P{2cm}|P{2cm}|P{2cm}|} 
			\hline
			\bf  F1-Score
			(\%)  & \bf Sensitivity
			(\%)
			& \bf Specificity
			(\%)
			& \bf  Precision
			(\%)
			& \bf Accuracy
			(\%)
			\\
			\hline
			89.35 & 91.9 & 86.2 & 86.94 & 89.05 \\
			\hline
		\end{tabular}
		\label{table:detection_pm}
		\vspace{-5pt}
	\end{table*}
	
	\begin{table}
		\caption{Normalized Confusion Matrix for Cough Detection (in Percentage).}
		\centering 
		\makegapedcells
		\begin{tabular}{cc|cc}
			\multicolumn{2}{c}{}
			&   \multicolumn{2}{c}{\bf Predicted} \\
			&       &   No Cough &   Cough              \\ 
			\cline{2-4}
			\multirow{2}{*}{\rotatebox[origin=c]{90}{\bf Actual}}
			& No Cough   & 86.2   & 13.8                 \\
			& Cough    & 8.1    & 91.9                \\ 
			\cline{2-4}
		\end{tabular}
		\label{table:detection_cm}
		\vspace{-7.5pt}
	\end{table}
	
	\section{Cough Detection System}
	
	The effectiveness of a \ac{CNN} typically scales with the size of the dataset used for training the model.
	Unfortunately, there is a particular scarcity of data in terms of cough audio. The results of prior works are difficult to compare with each other due to differing datasets used for model training. 
	Cough data from a small sample of individuals~\cite{Amoh2016} will accurately identify coughs among these individuals and those similar, but may not be generally applicable to a larger populace.
	Studies with a large number of subjects and thus cough samples~\cite{Aykanat2017,Sharan2019} lead to accurate detection models, but datasets of this size for cough audio are typically collected in a controlled environment such as a hospital.
	Since the purpose of our CNN is to distinguish coughs within clips of environment sounds, controlled recording environments do not suit our purpose.
	Coughs cropped from online sources~\cite{Pramono2016}, while numerous, vary wildly in audio quality.
	This variance, however, works towards the benefit of a more robust detection and diagnosis model that is able to identify relevant cough audio in less controlled conditions.
	Our paper takes a similar approach to acquiring the data used for model training.
	
	\subsection{Detection Database Description}
	
	The database used for training the \ac{CNN} in cough detection is composed of various modified audio clips gathered from open-source online sources~\cite{piczak2015environmental}.
	Each of these audio files originally contained at least one cough event and are cropped to a length of five seconds with the full cough contained at some arbitrary point within.
	Audio that contained coughs in more than a five second period are separated into multiple files.
	The exact number of coughs per file is left intentionally non-standardized.
	This database is balanced by a collection of environmental and speech sounds, comprised primarily of audio from the ESC-50 database.
	In addition to ESC-50, this non-cough half of the database is supplemented by sound clips taken from unused portions of the original cough audio that did not contain cough events.
	The final database contains $993$ cough samples and $993$ non-cough samples.
	
	
	
	For our detection database, the input raw audio clips are transformed into Mel-spectrogram, resulting in a 2-dimensional image where one dimension represents time, other dimension represents frequency and the value of pixels in the image represent the amplitude. 
	The resulting images of pixel size $432 \times 288 \times 3$ are then converted to gray-scale to unify the intensity scaling and are then compiled to form the final database of cough and no cough environmental audio clips. 
	
	\subsection{Detection CNN Structure}
	
	The relative success of using a \ac{CNN} for cough detection from recorded audio via image recognition serves as the basis for the machine learning model used for this system.
	Although our goal is to analyze a relatively larger time frame of five seconds, the model structure serves as a workable foundation.
	
	Overview of used \ac{CNN} structure is shown in \cref{fig:cnn}.
	Due to the high resolution of input image, we begin with a $2 \times 2$ max-pooling layer to lower input dimensions. 
	This will lower the required overall model complexity before proceeding. This is followed by a block of two convolutional layers, each having $32$ filters of size $5 \times 5$, a stride of $(1,1)$, ReLU activation function and a $2 \times 2$ max-pooling layer utilizing a $0.2$ dropout. The features learned from this convolutional block are flattened before passing them to two fully connected layers, each having $128$ neurons and ReLU activation function. These fully connected layers utilize a dropout of $0.5$. The final layer is the softmax classification layer having $2$ neurons to distinguish between cough and non-cough for the given input.
	
	The number of convolutional and fully connected layers are kept low to minimize potential overfitting issues. The database file count is increased to assist in reducing overfitting as well.
	After testing different activation functions, ReLU is considered because of its good performance in this case.
	ReLU is also the current standard for \acp{CNN}.
	Adam is used as the optimizer due to its relatively better efficiency and flexibility.
	A binary log loss function completes the detection model.
	
	\subsection{Detection Model Training and Results}
	
	The $1,986$ cough samples with $993$ cough items and $993$ non-cough items, are split into \SI{70}{\percent} training, \SI{15}{\percent} validation and \SI{15}{\percent} testing datasets. The model is then trained with a batch size of $32$ to find its optimal weights using an early stopping criteria on the validation dataset, so that the model training stops when it performs best on the validation dataset.
	Finally, the calculated performance metrics of the trained cough detection model on the testing dataset and the confusion matrix in \cref{table:detection_pm} and \cref{table:detection_cm}, respectively.

	These results reflect the performance of our system over a diverse dataset that comes from a comparatively huge swath of audio quality, cough volume and quantity.
	Much of the audio tested already contains ambient noise  due to the varied range of sources used for collecting the data. Due to the large file count, potential biases such as environmental consistencies in instances where coughs would typically occur are lessened, and will continue to diminish as the dataset increases in size. Hence, it is important to highlight that the performance of our proposed detection system is likely to improve further, for example, in controlled environments, which were typically the focus of prior works.
	
	%
	
	\begin{table*}
		\caption{Performance Metrics for Cough Diagnosis Classifier.}
		\centering
		\begin{tabular}{|P{1.9cm}|P{1.8cm}|P{1.9cm}|P{1.9cm}|P{1.8cm}|P{1.8cm}|} 
			\hline
			&	\bf  F1-Score
			(\%)  & \bf Sensitivity
			(\%)
			& \bf Specificity
			(\%)
			& \bf  Precision
			(\%)
			& \bf Accuracy
			(\%)
			\\
			\hline
			Overall&-&-&-&-& 89.60\\ \hline
			Pertussis&	94.43&	95.00&	96.90&	93.87	&-\\\hline
			Bronchitis&	85.74&	93.80&	87.50	&78.95	&-\\ \hline
			Bronchiolitis&		88.89	&80.00&	100.00&	100.00&	-\\ 
			\hline
		\end{tabular}
		\label{table:diagnosis_pm}
		\vspace{-5pt}
	\end{table*}
	
	\begin{table}[t]
		\caption{Normalized Confusion Matrix for Cough Diagnosis (in Percentage).}
		\centering 
		\makegapedcells
		\begin{tabular}{cc|ccc}
			\multicolumn{2}{c}{}
			&   \multicolumn{3}{c}{\bf Predicted} \\
			&       &   Bronchiolitis & Bronchitis & Pertussis \\ 
			\cline{2-5}
			\multirow{3}{*}{\rotatebox[origin=c]{90}{\bf Actual}}
			& Bronchiolitis   & 80.0 & 20.0 & 0.0                 \\
			& Bronchitis    & 0.0 & 93.8 & 6.2
			\\
			& Pertussis    & 0.0 & 5.0 & 95.0
			\\
			\cline{2-5}
		\end{tabular}
		\label{table:diagnosis_cm}
		\vspace{-5pt}
	\end{table}
	
	\section{Cough Diagnosis System}
	
	The data scarcity issue, as in the case of cough detection dataset, is also present for the diagnosis model, but compounded with the additional problem of labeled cough audio being considerably more scarce. Since the existing database for detection is comprised primarily of unlabeled cough audio with no identified illness relation, the same cannot be used as diagnosis training data. 
	To train our cough diagnosis  system,  we collected  cough sounds from 35 bronchiolitis, 131 pertussis and 102 bronchitis patients. 
	The performance and reliability of the diagnosis system is likely to improve as more data becomes available. 
	However, even with small training data, very promising results have been observed on unseen test samples, as reported later in diagnosis results section.
	
	\subsection{Diagnosis Database Composition}
	
	Since the purpose of the diagnosis model is not to detect the presence of any number of coughs but to categorize cough events according to their subtle differences, the input audio files for this model are much shorter in length. Hence, we adopt a different data preparation process for making it suitable for training.
	The sound files are first cut to a single cough event with the longest of these files lasting two seconds, the rest are buffered to the same length with noiseless audio to maintain a fixed two second length for all samples. This fixed length is set for the spectrogram conversion process that comes afterwards and all the two-second audio files are converted into $432 \times 288 \times 1$ Mel-spectrograms images. Finally, all images in the cough diagnosis dataset are then converted to gray-scale to unify its intensity scaling and reduce the size of each sample to facilitate the model training process.
	
	\subsection{Diagnosis CNN Structure}
	
	The classification model used for cough detection can also be applied to the labeled cough data originating from several different illnesses.
	However, due to the subtle differences between two coughs of differing illnesses, as compared to cough and non-cough event, a more complex set of layering is required for the  diagnosis model. 
	The structure of the \ac{CNN} used for diagnosis is similar to the detection \ac{CNN} structure in the beginning (depicted in \cref{fig:cnn}). However, after the $2 \times 2$ max-pooling layer at the end of convolutional block, another similar convolutional block comprising of two convolutional layers  is added before being passed on to a similar structure of two fully-connected layers with an identical neuron count of $128$, activation type of ReLU, and dropout of $0.5$. However, the final output layer now has $3$ neurons and Softmax activation function to classify the input between three possible diseases.
	
	\subsection{Diagnosis Model Training and Results}
	
	The diagnosis database is comprised of $268$ total cough sound items, distributed among bronchiolitis, pertussis, and bronchitis at counts of 35, 131, and 102, respectively.
	The database is again split into \SI{70}{\percent} training, \SI{15}{\percent} validation and \SI{15}{\percent} testing sets. The model is trained using an early stopping criteria on the validation dataset, so that the model training stops when it performs best on the validation dataset. Finally, the calculated performance metrics of the trained cough diagnosis model on the testing dataset and the confusion matrix are reported in \cref{table:diagnosis_pm} and \cref{table:diagnosis_cm}, respectively.
	
	
	
	
	
	These performance metrics clearly show the ability of this machine learning model to distinguish between multiple illnesses based on various cough sounds. It is also important to note that in previous works, the focus is usually to diagnose a specific illness from cough sounds, whereas we distinguish between multiple types of disease coughs in the same model.
	Due to the unbalanced nature of the database used for training the diagnosis model, F1-Score is more important here rather than the accuracy for making conclusions on the model performance.
	Since the data count is relatively low for the number of categories, more than one metric better interprets model performance.
	The potential bias mentioned for detection, environmental consistencies between disease categories unrelated to the cough audio structure, is also a relevant concern in this case.
	For example, individuals with a known case of bronchitis would typically be situated in hospitals more frequently than those with a flu or a cold. This issue would again be alleviated by a larger volume of data or perhaps a controlled variance in making sure that one category of data is not overtly reliant on background sounds for its unique model training identifiers. Regardless, the achieved performance, given the number of disease categories, is indicative of high level of success in diagnosing between specific illness-based coughs.

	\section{Discussion and Conclusion}
	
	We present a low complexity diagnostic tool for screening respiratory infections that utilizes a \ac{CNN}-based approach  to  detect  cough  within various types of environment noise and diagnose different potential illnesses based on their unique cough audio features. The approach uses cough audio and converts it to Mel-spectrograms for both detection and diagnosis. Experiments show that \ac{CNN} is capable of accomplishing these tasks with a high level of accuracy. This is achieved in spite of limited available cough data. With only a small number of modifications, the low-complexity network model can be trained sufficiently for both ends.
	
	The proposed model performance is likely to improve given a larger training dataset. It is also important to highlight that we did not perform exact comparisons of our results with the literature here. This is because of several reasons including difference in types and size of dataset or because of the black-box models. However, in the future work, we plan to extend our proposed model and compare it with the other models that have enough details available using our dataset.
	
	
	\section*{Acknowledgment}
	This work is supported by National Science Foundation under Grant Number 1559483. For more details, please visit www.AI4Networks.com.


	\bibliographystyle{IEEEtran}
	\balance
	\bibliography{references}


\end{document}